\documentclass[aoas]{imsart}

\arxiv{math.PR/0000000}
\newcommand{\xmin}{x_{\min}}

\usepackage{graphicx}
\usepackage{amssymb,amsfonts,amsmath,mathtools}
\usepackage{booktabs}
\usepackage{lipsum}
\usepackage{algorithm}
\usepackage{microtype, color, array}
\newcolumntype{L}[1]{>{\raggedright\let\newline\\\arraybackslash\hspace{0pt}}m{#1}}
\graphicspath{{graphics/}}
\RequirePackage{natbib}

\begin{document}

\begin{frontmatter}

  \title{Estimating the number of casualties in the American Indian war: a
    Bayesian analysis using the power law distribution} \runtitle{A Bayesian
    approach to the power law distribution}
%\thankstext{T1}{Footnote to the title with the `thankstext' command.}

\begin{aug}
  \author{\fnms{Colin S.}  \snm{Gillespie}\corref{}\ead[label=e1]{colin.gillespie@newcastle.ac.uk}}%
  \runauthor{Gillespie, CS}
  \affiliation{Newcastle University}
  \address{School of Mathematics \& Statistics, Newcastle upon Tyne\\
          \printead{e1}}
\end{aug}

\begin{abstract} The American Indian war lasted over one hundred years, and is a
  major event in the history of North America. As expected, since the war
  commenced in late eighteenth century, casualty records surrounding this
  conflict contain numerous sources of error, such as rounding and counting.
  Additionally, while major battles such as the Battle of the Little Bighorn
  were recorded, many smaller skirmishes were completely omitted from the
  records. Over the last few decades, it has been observed that the number of
  casualties in major conflicts follows a power law distribution. This paper
  places this observation within the Bayesian paradigm, enabling modelling of
  different error sources, allowing inferences to be made about the overall
  casualty numbers in the American Indian war.
\end{abstract}
\end{frontmatter}

\section{Introduction} \label{s:intro}

The American Indian war spanned the time period, 1778 to around 1890, and
covered a wide geographical area. As such, the participants, and to a certain
extent, technology changed throughout the war. Although some authors have
attempted to divide the period into separate conflicts (see, for example,
\cite{Clodfelter2008} and \cite{Axelrod1993}), there is no agreed time period
division. This time was characterised by low-level violence with occasional
large scale battles, such as, the 1794 Battle of Fallen Timbers, 1811 Battle of
Tippecanoe. and 1876 the battle of the little Bighorn.

Around the time of American independence ($\sim\!\!1780$) the majority of the
population lived near the sea and the country had a limited military capacity.
Consequently most of the casualties were due to small skirmishes which
occasionally led to larger conflicts.

During the early nineteenth century, many tribes were relocated to lands west of
the Mississippi river. Most removals were met with relatively little violence,
but in a few cases, tribes fought long conflicts to stay on their land. The
relocations carried on throughout the nineteenth century, with native Americans
being increasingly confined to reservations. 

It has been observed that the severity of many violent events follow a power law
distribution. In a seminal paper, \cite{Richardson1948} divided international
and domestic instances of violence between $1820$ and $1945$ into logarithmic
groups. A subsequent paper \citep{Richardson1960}, with an updated data set,
demonstrated that the frequency of entries in each logarithmic group, followed a
simple multiplicative law: for each $10$--fold increase in the number of
casualties, the frequency decreased by around a factor of three.
\cite{Cederman2003} updated this analysis with data (restricted to interstate
wars) from the Correlates of War (COW) Project \citep{Geller1998}.
\citeauthor{Cederman2003} found that the multiplicative law seemed to hold for
the last two centuries. He then developed agent-based models to suggest possible
generative mechanisms for this ``law''.

\cite{Clauset2007} extended this work to include terrorist attacks since 1968.
They noted that the frequency-severity statistics of terrorist events are scale
invariant and concluded that there is no fundamental difference between small
and large attacks. Similarly, \cite{Bohorquez2009} investigated the severity of
insurgent attacks within nine separate conflicts and again found that the data
had a power law structure. 

Power law distributions are often described as ``scale-free'', indicating that
common small events are qualitatively similar to large rare events. If this
pattern can be detected in empirical data, it may indicate the presence of an
interesting underlying process. Being able to detect whether a system does or
does not follow a power law can provide hints to the generative mechanisms at
work. During the twentieth century, numerous examples of these heavy tailed
distributions have been used to describe a variety of different phenomena,
including city size, word frequency and the productivity of scientists; see, for
example \cite{Newman2005}, \cite{Dewez2013} and \cite{Bell2012}.

This apparent ubiquity of power laws in a wide range of disciplines was
questioned by \cite{Stumpf2012}. The authors pointed out that many ``observed''
power law relationships are highly suspect. In particular, estimating the power
law exponent from a log-log plot, whilst simple and appealing, is a very poor
technique for fitting these types of models. Instead, a systematic, principled
and statistically rigorous approach should be applied, such as those by
\cite{Clauset2009}, \cite{Breiman1990}, \cite{Dekkers1993}, and
\cite{Drees1998}.

The common problem with fitting the power law distribution is that analyses
typically assume that the power law phenomena is present in the \textit{tail} of
the distribution. This results in the need to estimate both the scaling
parameter $\alpha$ and the lower bound threshold for where the power law
behaviour begins, $\xmin$. \cite{Clauset2009} (CSN) introduced a principled set
of methods for fitting and testing power law distributions. Their approach is
straightforward and couples a distance-based test for estimating $\xmin$ with
estimation of $\alpha$ via maximum likelihood. Alternative models, such as the
log normal distribution, can be compared using a likelihood ratio test
\citep{Vuong1989}. However this method does have a few issues. First, after
estimating $\xmin$, we \textit{discard} all data below this threshold. Second,
it is unclear how to compare distributions with different $\xmin$ values. Third,
although it is possible to make predictions in the tail of the distribution,
making future predictions over the entire data space is not possible since
values less than $\xmin$ have not been directly modelled \citep{Clauset2013}.

One further difficulty with the method proposed by CSN is that it is not
straightforward to incorporate an error model during the inference stage.
Recently, \cite{Virkar2014} considered a model for binned data in which the
observed data has been rounded or grouped. Essentially, the authors propose a
modification to the CSN algorithm. However, this modification is difficult to
generalise to other error structures and also suffers from the same issues as
the original method.

In this paper we use a power law distribution to model the number of casualties
sustained in the American Indian war. By adopting a Bayesian approach we are
able to model the under-reporting of casualties, incorporate prior beliefs on
the parameters and provide predictions for the true number of casualties. In the
following section, we discuss the power law distribution and the CSN method in
more detail. In section 3, we give a brief introduction to the American Indian
war, before moving on to a fully Bayesian analysis of the model. The paper
closes with a discussion in section 4.

\section{The power law distribution}

The discrete power law distribution has the probability mass function (PMF)
\begin{equation}\label{2}
  \text{Pr}(X=x) = \frac{x^{-\alpha}}{\zeta(\alpha, \xmin)}, \quad x = \xmin, \xmin+1, \ldots
\end{equation}
where
\begin{equation}\label{3}
\zeta(\alpha, \xmin) = \sum_{n=0}^{\infty} (n + \xmin)^{-\alpha}
\end{equation}
is the generalized zeta function which converges provided $\alpha > 1$
\citep{Abramowitz1970}. When $\xmin=1$, $\zeta(\alpha, 1)$ simplifies to the
standard zeta function, $\zeta(\alpha) = \sum_{n=1}^{\infty}
n^{-\alpha}.$

The value of $\alpha$ determines which moments diverge. For example, if $1<
\alpha \le 2$, all moments diverge, i.e., $E[X^m] = \infty$, $m=1, \ldots$; if
$2 < \alpha \le 3$, all second and higher-order moments diverge, i.e., $E[X^m] =
\infty$, $m=2, \ldots$; and so on.

\subsection{Parameter inference}

\cite{Clauset2009} show that when $\xmin$ is known, an approximate maximum
likelihood estimate for the discrete power law scaling parameter $\alpha$ is
\begin{equation}\label{5}
  \hat \alpha \simeq 1 + n \left[\sum_{i=1}^n \ln \left(\frac{x_i}{\xmin - 0.5}\right)\right]^{-1} \;.
\end{equation}
The likelihood estimate for the continuous power law is almost identical, but
with the $0.5$ removed from the denominator. In many practical situations, it is
argued that only the tail of the distribution follows a power law, but the value
of $\xmin$ is unknown. Unfortunately, as the value of $\xmin$ increases, the
amount of data that is \textit{discarded} also increases, so it is clear that
some care must be taken when estimating this parameter. Estimation via maximum
likelihood is not appropriate since for each value of $\xmin$, the likelihood
function is calculated using a different data set.

Until recently, a common approach used to estimate $\xmin$ has been from a
visual inspection of the data on a log-log plot. Clearly, this is error prone
and subjective \citep{Stumpf2012}. The \textit{de-facto} method for estimating
the lower bound $\xmin$ and corresponding scaling parameter $\alpha$ is the
Kolmogorov-Smirnov technique proposed by \cite{Clauset2009}. Denoting $F(x)$ and
$\hat F(x)$ to be the CDFs of the model and data respectively (for $x \ge
\xmin$), the lower bound $\xmin$ can be estimated by minimising the statistic
\begin{equation}
  D = \max_{x \ge \xmin} \vert F(x) - \hat F(x) \vert \;.
\end{equation}
For a given value of $\xmin$, the MLE standard error of $\alpha$ can be derived
analytically. However, this ignores the additional uncertainty due to the lower
bound estimate, $\xmin$. To quantify uncertainty a bootstrap procedure can be
used \citep{Efron1993}. Essentially, we sample with replacement from the
original data set and then re-infer the parameters at each step using
Algorithm~\ref{A1}.

\begin{algorithm}[t]
  \caption{Estimating the uncertainty in $\xmin$ \citep{Clauset2007}}\label{A1}
 \begin{tabular}{@{}ll@{}}
   {\small 1:} & Set $N$ equal to the number of values in the original data set. \\
   {\small 2:} & \textbf{for} \texttt{i} in \texttt{1:B}:\\
   {\small 3:} & $\quad$ Sample $N$ values (with replacement) from the original data set. \\
   {\small 4:} & $\quad$ Estimate $\xmin$ and $\alpha$ by minimising the Kolmogorov-Smirnov statistic.\\
   {\small 5:} & \textbf{end for} \\
  \end{tabular}
\end{algorithm}

To test whether the data set of interest follows a power law, we can employ
another bootstrapping procedure. Essentially, for each bootstrap we simulate a
new data set using the inferred parameters and refit the model. However, this
can be computationally prohibitive.

The Kolmogorov-Smirnoff procedure developed by CSN is principled, relatively
straightforward to apply and is a substantial improvement over estimating the
parameters by eye. Consequently it is widely used and has been implemented in
both python \citep{Alstott2014} and R \citep{Gillespie2015} programming
languages.

Many data sets are collected with error. Thus even if $X$ follows a power law,
we typically observe a corrupted version. For example, in \cite{Virkar2014}, the
authors extend the CSN method to fit \textit{binned} data, that is, within a set
of $k$ boundaries, $0 < b_1 < b_2 < \ldots < b_k$, we only observe the number of
observations that fall within a particular region. So as $k \rightarrow \infty$,
we fully observe the process. Again the difficulty with the estimation process
is that we are only modelling the tail of the distribution and so need to infer
the cut-off $\xmin$. Extending this framework to deal with more complex error
models is non-trivial. For example, in this paper we study the American-Indian
data set where we observe $X$ with probability $p$ and a binned value with
probability $1-p$, i.e. a proportion of the data has been binned/rounded.

\section{Modelling casualties in the American Indian war}

\begin{figure}[t]
  \centering 
  \includegraphics[width=0.85\textwidth]{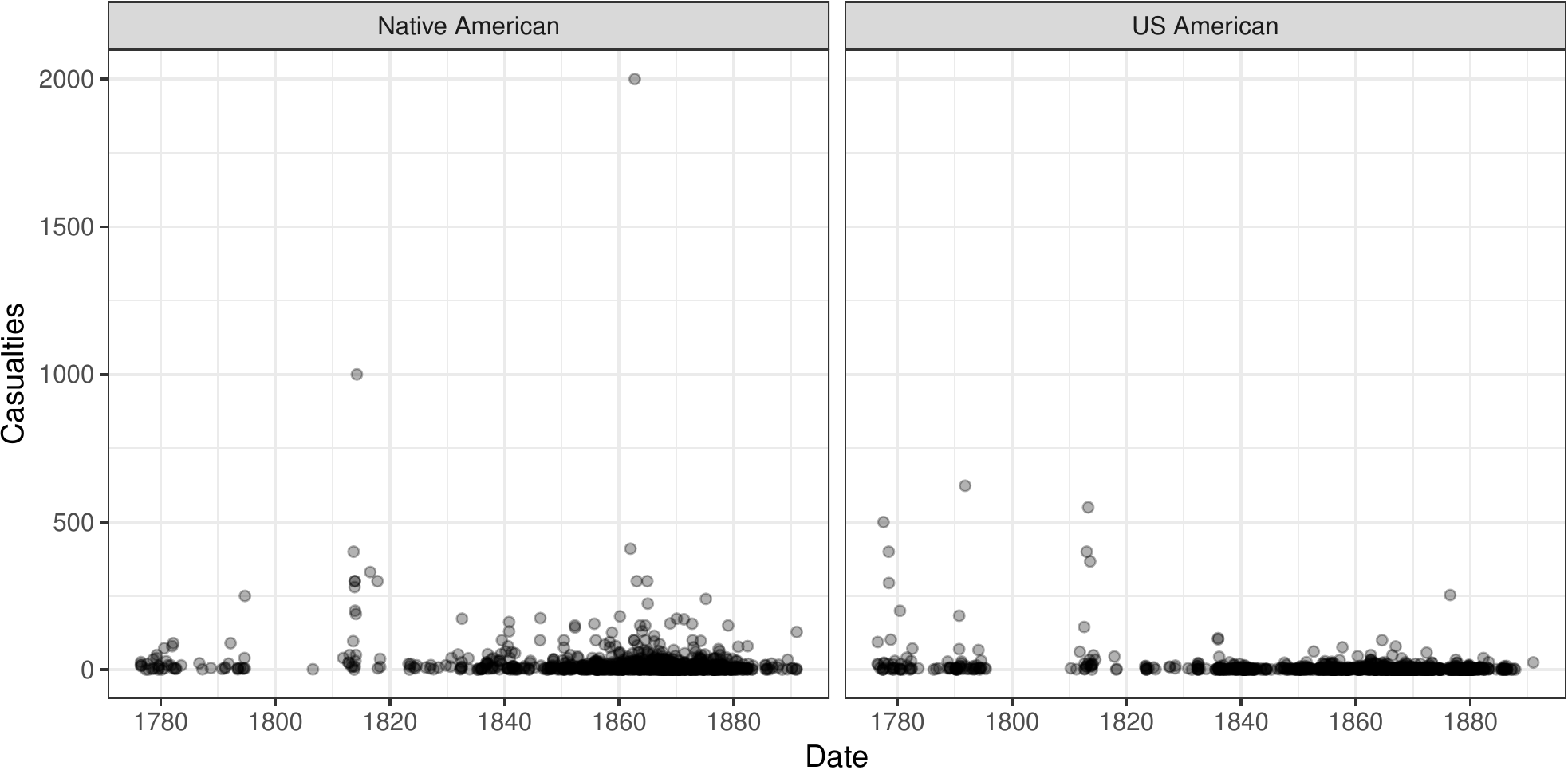}
  \includegraphics[width=0.85\textwidth]{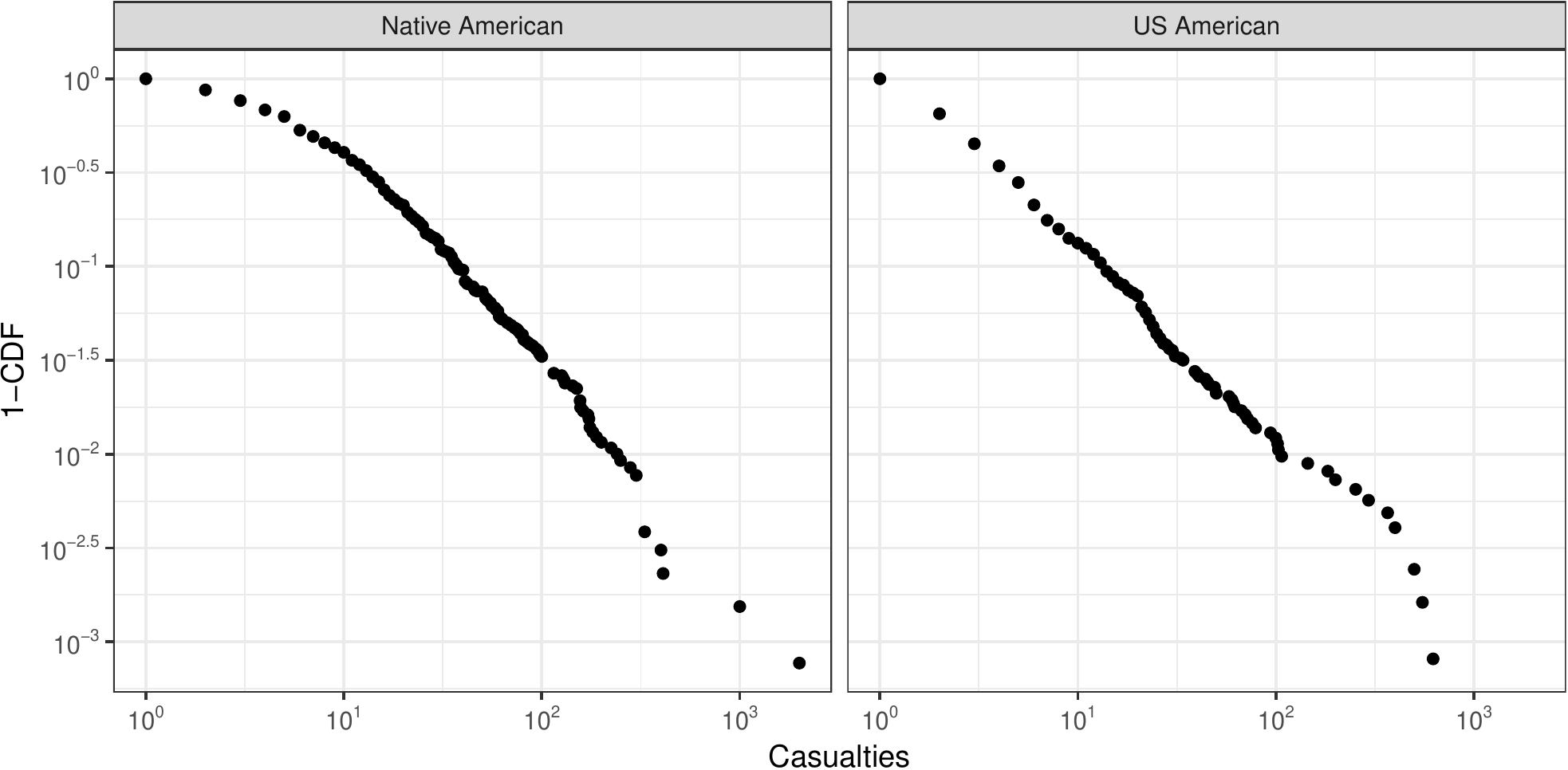}
  \caption{Casualities sustained by the Native American and US American sides
    during the American Indian war, 1776--1890. Each point represents a
    particular conflict. Data originally collected and analysed by
    \cite{Friedman2013}. (b) Empirical CDFs of the casualty data set.}\label{F1}
\end{figure}

This paper builds on \cite{Friedman2013}, by attempting to infer the number of
casualties that occurred during the American Indian war. Figure \ref{F1}a shows
the casualties sustained by both sides. Small scale conflicts are prominent for
both sides during the war. For example, in over 50\% of the US American
conflicts there were only one or two recorded casualties. For the Native
Americans, this proportion was around 25\%. Figure \ref{F1}b gives the $1-$ the
empirical CDFs of the data set, where each point represents a specific battle.
For power law distributions, the points lie on a straight line. Clearly, for
both combatants the points only lie on a line in the tail of the distribution.

Unsurprisingly it is extremely unlikely that data collection alone can give us a
precise estimate of the number of casualties sustained by both sides from
historical conflicts. The primary issue with the data set is under reporting,
particularly with the native American casualties. A secondary issue is data
quality. In addition to the usual mis-counting in both the native and US
records, there are clear rounding effects in the data.

\subsection{The underlying process}

A \textbf{casualty} is defined as a person captured, mortally wounded, or killed
in a \textit{particular} battle or skirmish. Casualties include military
engagements that occurred within the continental United States and also any
pursuits into neighbouring territories.

In our study of the American Indian war, we assume that the generative mechanism
for the numbers of casualties in a particular battle or skirmish for the Native
Americans follows a power law distribution, resulting in a likelihood
contribution
\begin{equation}\label{7}
  \text{Pr}(W_{i, N} = w_{i,N}) = \frac{w_{i, N}^{-\alpha_N}}{\zeta(\alpha_N)}
  \quad i=1, \ldots, n_{true, N}
\end{equation}
where $\zeta(\cdot)$ is the standard zeta function, $w_{i, N}$ is the true
number of casualties sustained by the Native Americans in battle/skirmish $i$,
and $n_{true, N}$ is the true number of battles/skirmishes for the Native
Americans. Note that the total number of casualties sustained by the Native
Americans is given by $\sum_{i=1}^{n_{true, N}} w_{i, N}$. Similarly for the US
forces we have
\[
 \text{Pr}(W_{i, U} = w_{i,U}) = \frac{w_{i, U}^{-\alpha_U}}{\zeta(\alpha_U)}
  \quad i=1, \ldots, n_{true, U} \;.
\]
Ideally we would have joint records for the casualties sustained by US and
Native American forces for each battle. This would enable us to jointly model
the number of casualties for each side. Unfortunately the data do not contain
this information since battles are missing and casualties are recorded (if at
all) after the event. Instead we link the forces by having an informative joint
prior on $(\alpha_N, \alpha_U)$; this is discussed in more detail in section
\ref{Bayesian}.

To make the notation clearer in the following discussion, we will drop the
subscripts $N$ and $U$ from the random variables and parameters.

The idea that the number of casualties occurring in a battle comes from a
distribution where the variance and/or mean are infinite is not plausible: for
any given conflict there is maximum number of casualties that can be sustained.
However, it does provide a mechanism for characterising the underlying
distribution; this assumption is investigated in section \ref{modelfit}.

\begin{figure}[t]
  \centering
  \includegraphics[width=0.6\textwidth]{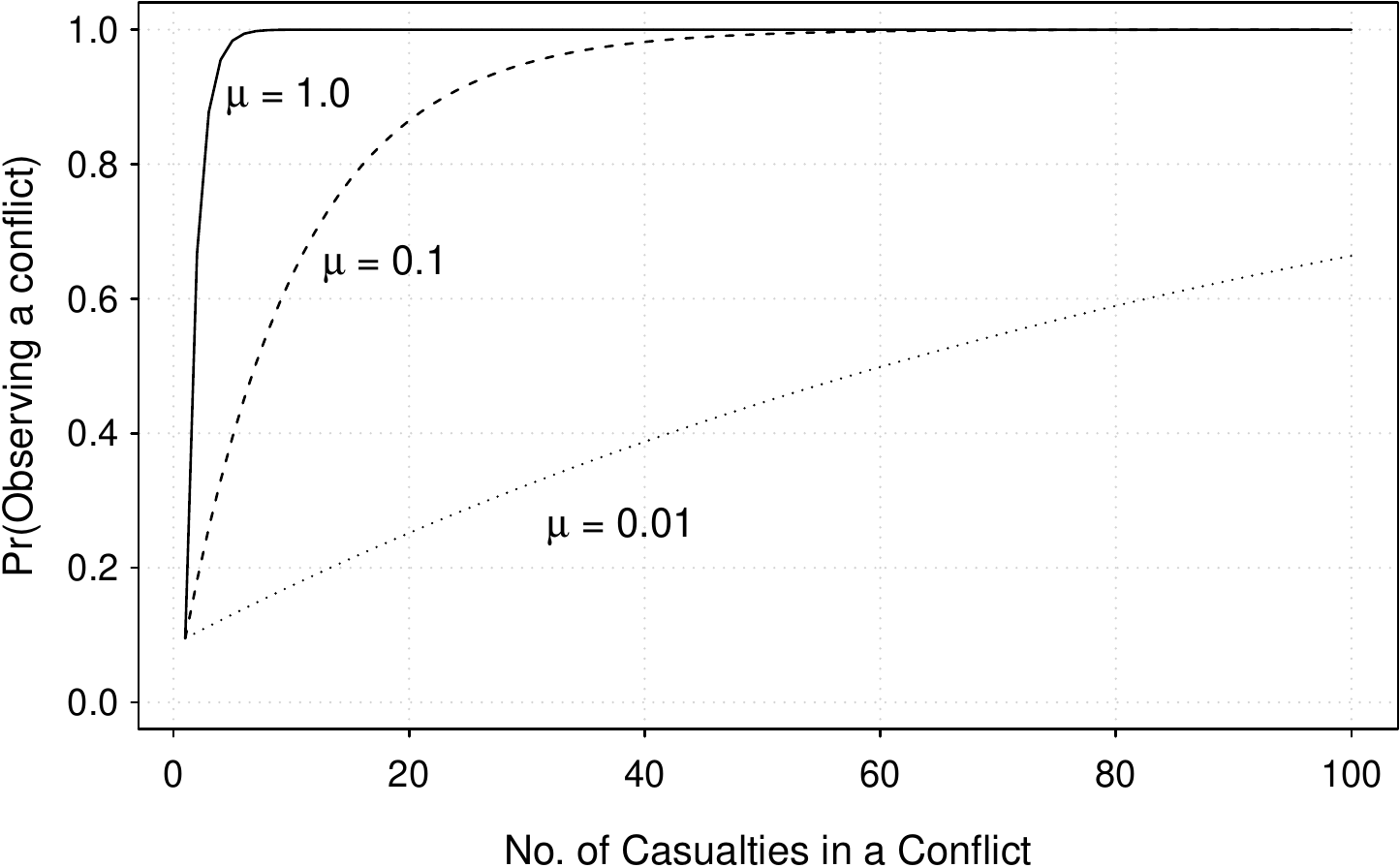}
  \caption{The probability of observing an event for $\lambda=0.1$ and $\mu$ as
    indicated. The probability of observing an event where there is a single
    casualty, is $1-e^{-0.1} \simeq 0.1$.}\label{F2}
\end{figure}

Clearly historical records are not perfect and some conflicts will be omitted.
However, battles are not missing at random. Instead conflicts that sustain only
a small number of casualties are more likely to be omitted than large scale
conflicts. For example, it is unlikely there was a conflict where US forces
sustained over $100$ casualties that wasn't recorded. To model the probability
that a conflict was omitted we use a logistic-type function, with $i=1, \ldots,
n_{true}$,
\begin{equation}\label{8}
  \text{Pr}(X_i = w_i \vert W_i = w_i) = 1 - \exp\{-\lambda - \mu (w_i-1)\}
  \quad \lambda, \mu > 0 
\end{equation}
otherwise $X_i$ is missing. The number of events that we observe is $n_{obs}$.
As the number of casualties, $w_i$ increases, the probability of observing and
recording a conflict, tends to one. Furthermore, the probability of observing a
conflict of size $w_i=1$ is $1-e^{-\lambda} \simeq \lambda$. A plot of the
missingness probability function (\ref{8}) is shown in Figure \ref{F2} for
different values of $\mu$, with $\lambda=0.1$. As $\mu$ increases, the
probability of observing an event also increases.

\subsection*{Data quality}

So far we have assumed that the underlying generative model is a power law, with
the probability of observing a battle following expression (\ref{8}). However
even when a battle has been recorded, it is likely that the record contains errors.

The first recording error we consider is a counting error. Since the size of the
error is likely to increase with the number of casualties, we will model this
error using a Poisson distribution as with this distribution, the variance
equals the mean. However, since missing observations are already captured using
expression (\ref{8}), we use a truncated Poisson distribution. Letting $Y_i$ to
be a noisy measurement of the true number of casualties in observed conflict
$i$, we have
\begin{equation}\label{9}
  \text{Pr}(Y_i = y_i \vert X_i = x_i) = \frac{x_i^{y_i}
    e^{-x_i}}{y_i!(1-e^{-x_i})}, \quad y_i=1, 2, \ldots, i=1, 2, \ldots, n_{obs}.
\end{equation}

\begin{table}[t]
\centering
\caption{Observed number of casualties for the US and Native
    American forces, e.g. there were $430$ conflicts where the number of US
    casualties was $1$. The data indicates that rounding to the nearest $5$ casualties has occurred.}\label{T1}
\begin{tabular}{@{}l lllll lllll@{}}
  \toprule
  &  \multicolumn{10}{l}{No. of Casualities} \\
  \cmidrule(l){2-11}
  & $1$ & $2$ &   $3$ &  $4$ &  $5$ &  $6$ & $7$ & $8$ & $9$ & $10$ \\
  \midrule
  US & $430$ & $247$ & $132$ & $78$ & $\mathbf{\phantom{0}83}$ & 
  $45$ & $22$ & $21$  & $10$ & $\mathbf{10}$ \\
  Native & $166$ & $139$ & $107$ & $69$ & $\mathbf{126}$ & $51$ & 
  $48$ & $34$ & $31$ &$\mathbf{49}$\\
  \bottomrule
\end{tabular}
\end{table}

\noindent A further source of error that is present is rounding or heaping
\citep{Crawford2015}. From Table \ref{T1} there is clear evidence that the
number casualties have been rounded to the nearest five: for the Native
American's there were $69$ and $51$ events where the number of casualties was
$4$ and $6$ respectively, whereas there were $126$ recorded conflicts where the
number of casualties was $5$. As might be expected, rounding seems to more
prevalent for the Native American casualty figures. For each battle, $i=1,
\ldots, n_{obs}$, we assume that no rounding occurs when $y_i =1 $ or $2$. For
values of $y_i > 2$, the rounding mechanism is modelled as
\begin{equation}\label{10}
  \text{Pr}(Z_i = z_i \vert Y_i = y_i) = 
\begin{cases}
y_i & \mbox{with probability } 1-p \\
5 \times \left(\left[\frac{y_i-2.5}{5}\right] + 1\right) & \mbox{with probability } p,
\end{cases}
\end{equation}
where $[\cdot]$ denotes the integer part.

A summary of the modelling process is given in Table \ref{T1b}.

\begin{table}[!b]
  \centering
\caption{A summary of the different modelling stages. The formal definitions
  are given at the indicated equation numbers.}\label{T1b}
  \begin{tabular}{@{}lL{8cm}l@{}}
    \toprule
    Variable & Description & Definition \\
    \midrule
    $W_i$ & The true number of casualties that occurred in conflict $i$. & (\ref{7})\\
    $X_i$ & Was the conflict recorded. & (\ref{8})\\
    $Y_i$ & The number of casualties in a recorded conflict $i$, with Poisson counting errors. & (\ref{9})\\
    $Z_i$ &  The observed historical value. The number of casualties in a recorded
    conflict $i$, with Poisson counting  and rounding/heaping errors.
    & (\ref{10})\\
    $n_{obs}$ & The total number of observed conflicts. \\
    $n_{true}$ & The total of number of conflicts (including missing battles).\\
    \bottomrule
\end{tabular}

\end{table}

\subsection{Bayesian parameter estimation}\label{Bayesian}

The inference task is two-fold. First, we wish to make statistically valid
statements about the unknown model parameters $(\alpha, \lambda, \mu, p)$.
Second, we wish to predict the true, unobserved, number of casualties sustained
by each force during the conflict. The Bayesian statistical inference approach
combines information from the data with prior parameter information. The
resulting posterior distribution enables us to make predictions about the actual
casualty rates.

We denote $\theta = (\alpha_U, \lambda_U, \mu_U, p_U, \alpha_N, \lambda_N,
\mu_N, p_N)$ to be the model parameters of both datasets and $z$ to be the
combined datasets, $z = (z_U, z_N)$ where the subscripts $U$ and $N$ denote the
US and Native American forces, respectively.

Prior distributions for the observation rates were obtained from the following:
\begin{itemize}
\item A reasonable lower bound for observing a conflict of size $1$, is to only
  observe $1$ in every one thousand battles, i.e.
  $1 - e^{-\lambda} \simeq 0.001$.
\item It is unlikely that we would record all conflicts of size $1$. Instead, we
  would expect to observe at most $95$\% of such events.
\item Casualties for the US forces were more likely to be recorded than the
  Native American forces.
\item It is unlikely that large scale conflicts were omitted.
\end{itemize}
This prior information is captured using a fairly weak bivariate log normal
prior, namely
\[
\left(
  \begin{matrix}
    \lambda_U \\
    \lambda_N
  \end{matrix}
\right)
\sim LN_2\left( 
  \left(
    \begin{matrix}
      \phantom{-}0.0 \\
      -3.0
    \end{matrix}
  \right), \,
  \left(
    \begin{matrix}
      1.0 & 0.6  \\
      0.6 & 2.0
    \end{matrix}
  \right)
\right)\;.
\]
The same (independent) prior was used for $(\mu_U, \mu_N)^T$.

For the power law parameters $\alpha_U$ and $\alpha_N$, we use independent
$U(1.5, 3)$ distributions. These end points were chosen as when $\alpha > 3$, we
are unlikely to have any large scale conflicts, while when $\alpha < 1.9$, the
power law generates values much larger than is feasible.

For the remaining heaping parameters ($p_U$ and $p_N$) we assume relatively weak
but proper prior specifications, namely, independent $U(0, 1)$ distributions. It
is worth noting that these priors could be made more informative. For example,
we might expect more rounding of the casualty figures for the Native American
forces, possibly leading to different priors for each force. However a
sensitivity analysis reveals that the posterior is relatively insensitive to
modest changes in these priors.

Therefore the posterior distribution for the parameters is
\begin{multline}
  \pi(\theta \,\vert\, z) \propto 
\pi(\lambda_U, \lambda_N) \pi(\mu_U, \mu_N) \times\\
\prod_{j=U,N} \pi(\alpha_j) \pi(p_j) \pi(w_j\vert \alpha_j)
\pi(x_j \vert w_j, \mu_j, \lambda_j) \pi(y_j \vert x_j) \pi(z_j \vert y_j, p_j).
\end{multline}

We use a Markov chain Monte Carlo (MCMC) algorithm to sample from the posterior
distribution. The parameter space was explored using a multivariate Gaussian
random walk proposal, with the tuning parameters obtained from a pilot run.

We could construct an MCMC sampler and propose the latent states $w_i$, $x_i$
and $y_i$. However, building efficient transition kernels for the latent states
is difficult since the data is both discrete and covers many orders of
magnitude. We can neatly circumvent this issue by directly integrating out the
latent state $y_i$, since
\[
\text{Pr}(z_i \vert x_i) = 
\begin{cases}
  f(z_i\vert x_i) & z_i = 1, 2, \\
  f(z_i\vert x_i)(1-p) + p\sum_{k=-2}^2 f(z_i-k\vert x_i) & z=5, 10, \ldots,\\
  f(z_i\vert x_i)  (1-p) & \text{otherwise}, \\
\end{cases}
\]
where $f(z_i\vert x_i)$ is the truncated Poisson distribution defined in
(\ref{9}).

Typically when we have a latent variable, such as $w_i$, we use an MCMC step to
propose missing values. However, another substantial computational saving can be
made by noting that we can integrate out uncertainty for $w_i$, since
\[
\text{Pr}(X_i = x_i) \propto \left(1 - \exp\left\{-\lambda - \mu (x_i-1) \right\}\right)
\times \frac{x_i^{-\alpha}}{\zeta(\alpha)}
\]
i.e. we do not propose unobserved battles. The posterior distribution for the
true number of battles, $n_{true}$ can be obtained \textit{post} MCMC by using
the posterior sample.

Proposing latent states $x_i$ directly via an independence sampler, such as a
power law distribution, resulted in a very low acceptance rate. Therefore we
used a random walk on the latent structure. At each iteration of the algorithm
ten $x$-values were selected at random and perturbed using the truncated Poisson
(TP) distribution (equation \ref{9}), i.e.
\[
x_i^* \vert x_i \sim TP(x_i)
\]
where the subscript $i$ refers to a particular value of $x$.

The proposed parameter values $\theta^*$ are accepted with probability
\begin{equation}
  \min\left\{1, 
    \frac{\pi(\theta^*)}{\pi(\theta)} 
    \times \frac{\pi(z_i^* \vert x_i^*, \theta^*)\pi(x_i^* \vert \theta^*)}{\pi(z_i
      \vert x_i, \theta) \pi(x_i\vert \theta)} 
    \times \frac{q(x_i \vert x_i^* )q(\theta \vert \theta^* )}{q(x_i^* \vert x_i)q(\theta^* \vert \theta)}
  \right\} 
\end{equation}
where $q(\cdot)$ is the multivariate Gaussian or truncated Poisson transition
kernel as appropriate.

\subsection{Simulation study}\label{Simulation}

The performance of the algorithm was examined by considering a simulated data
set for a single side. To mirror the real dataset, we set the power law scaling
parameter to $\alpha = 2.2$ and the parameters governing the probability of
observing an event at $\lambda = 0.007$ and $\mu = 0.05$. The probability of
rounding was set at $p=0.19$. Setting the total number of events to be
$n_{true} = 20,000$ in this simulation study, gave approximately
\[
\sum_{i=1}^{n_{true}} w_i \simeq 64,000 \quad \text{and}\quad \sum_{i=1}^{n_{obs}} z_i \simeq 31,000
\]
casualties. For simplicity, this simulation study only considers a single side,
hence we have a single $n_{true}$ value.

\begin{figure}[t]
  \centering 
  \includegraphics[width=\textwidth]{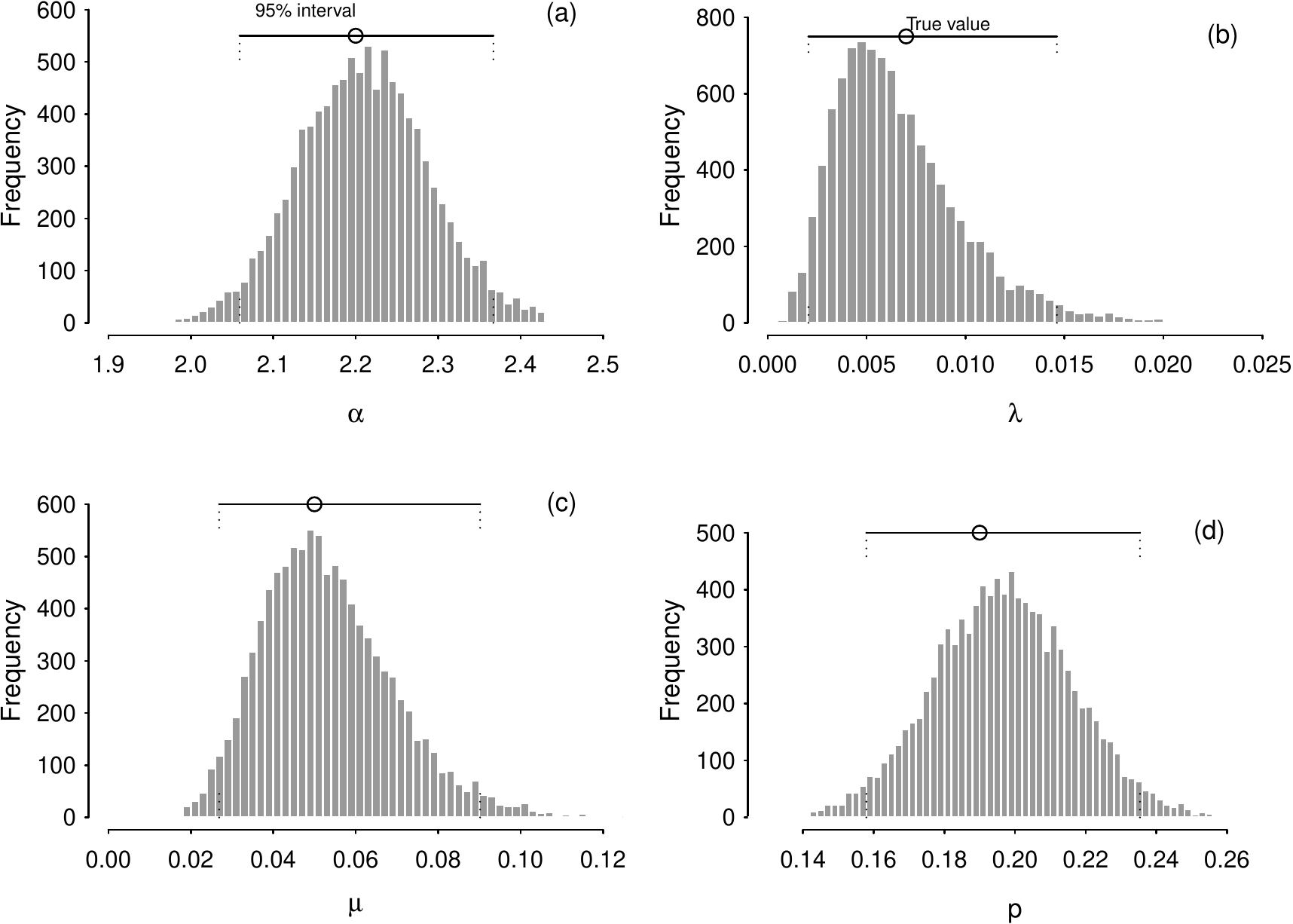}
  \caption{Marginal posterior distributions for the four model parameters and
    the predictive distributions for the total and observed number of casualties
    for the simulation study. For each marginal distribution, a 95\% prediction
    interval and the true value is shown. In all cases, the true value is within
    the posterior distribution.}\label{F3}
\end{figure}

\begin{figure}[t]
  \centering 
  \includegraphics[width=0.5\textwidth]{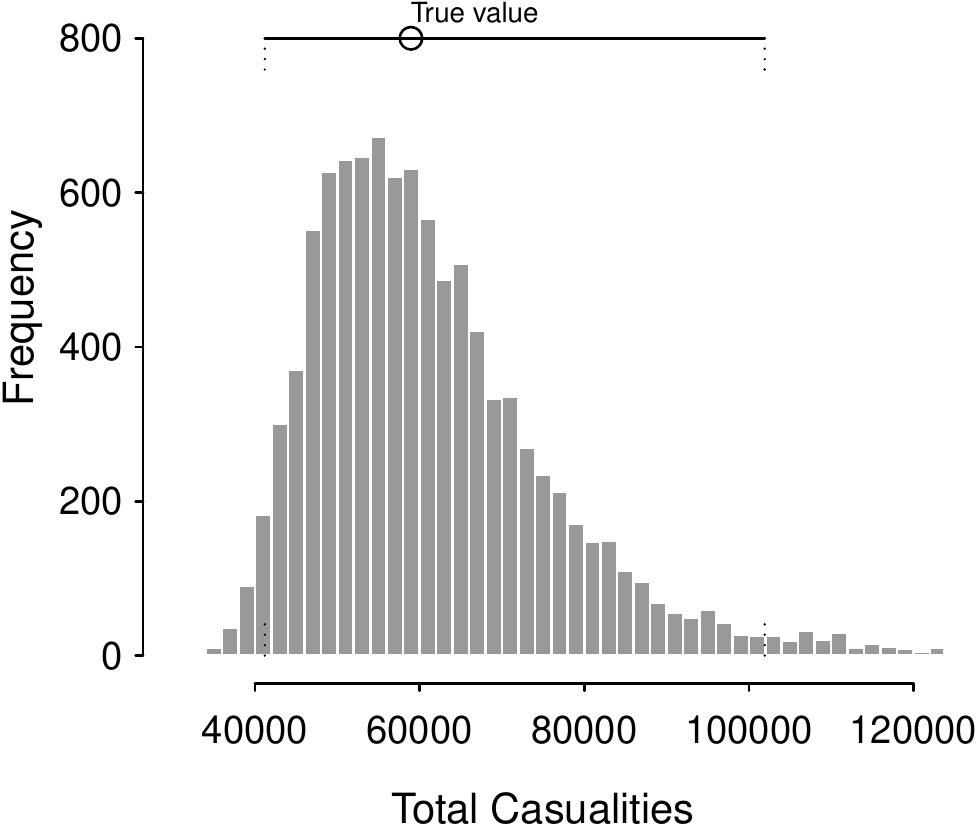}
  \caption{The predictive distribution for the total number of casualties in the
    simulation study.}\label{F4}
\end{figure}

After a pilot run to estimate the random walk tuning parameters, we ran the MCMC
algorithm (described in Section \ref{Bayesian}), for $1.1$ million iterations.
The first $100,000$ iterations were removed as burn-in and the remainder thinned
by a factor of $100$ iterations. This yielded a sample of $10,000$ iterates with
low auto-correlation to be used as the main monitoring run. See the
supplementary material for additional diagnostic plots (\cite{Gillespie2017}).

The posterior densities for the four parameters are given in Figure \ref{F3},
a--d. The true value is within the 95\% credible region for each parameter.

Integrating over parameter uncertainty, we obtain a posterior predictive
distribution for the true number of casualties (figure \ref{F4}). As might be
expected, there is considerable uncertainty in the total number of casualties.
In particular, there is a relatively long tail which is a result of the
underlying power law distribution. However, the true value is still within the
95\% credible region.

\subsection{Application to the American Indian war}

\begin{figure}[t]
  \centering 
  \includegraphics[]{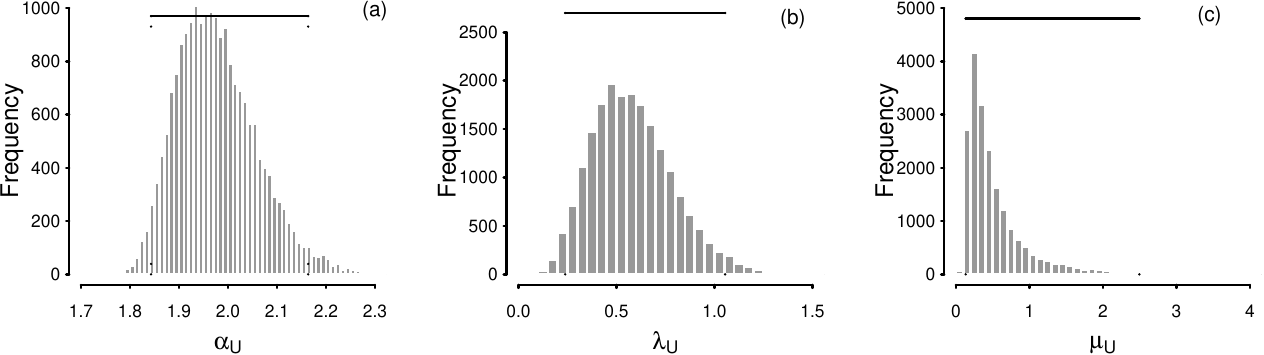}
  \includegraphics[]{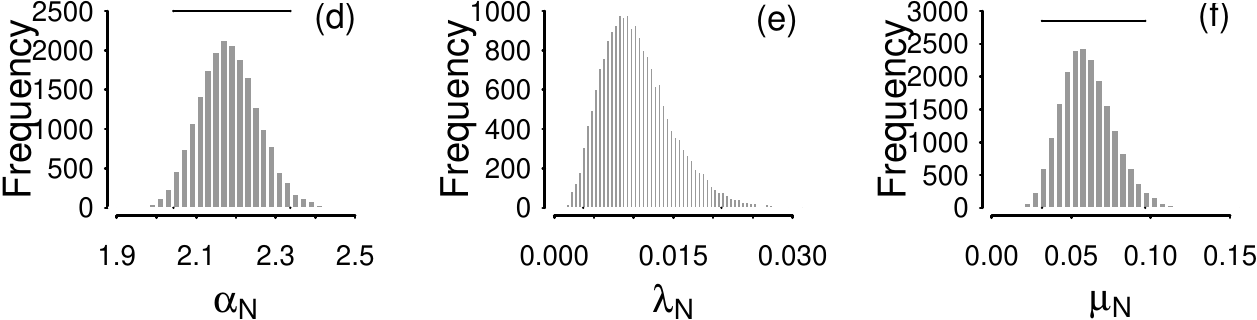}
  \includegraphics[]{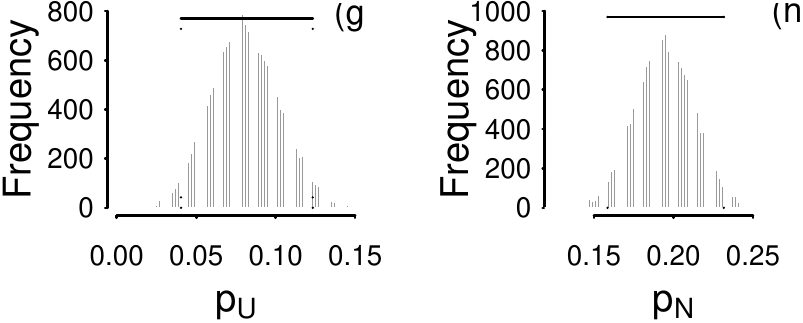}
  \caption{Marginal posterior distributions for the model parameters. For each
    distribution, a 95\% prediction interval. The subscripts $U$ and $N$ denote
    the US and Native American forces respectively.}\label{F5}
\end{figure}

Similar to the simulation study, a pilot MCMC run for each force was used to
estimate sensible transition kernel tuning parameters. The overall acceptance
rate was around $9\%$. The parameters $\alpha$, $\lambda$ and $\mu$ are highly
correlated, with a pairwise correlation coefficient of $r \simeq 0.8$. We ran
the MCMC algorithm for $2.1$ million iterations; discarding the first $100,000$
iterations and thinning the remainder by a factor of $100$. The total simulation
time was around $9$ hours. See the supplementary material for additional
diagnostic plots (\cite{Gillespie2017}).

The results for the US and Native American data sets, denoted with sub-scripts
$U$ and $N$ respectively, are given in figure \ref{F5}. For both data sets, the
power law scaling parameter is around $\alpha \simeq 2.1$, with $\alpha_N <
\alpha_U$. Intuitively, this makes sense since most casualties on each side
would be low level, and the maximum number of casualties in a single event would
be approximately the same for both sides.
 
\begin{figure}[t]
  \centering
  \includegraphics[width=0.85\textwidth]{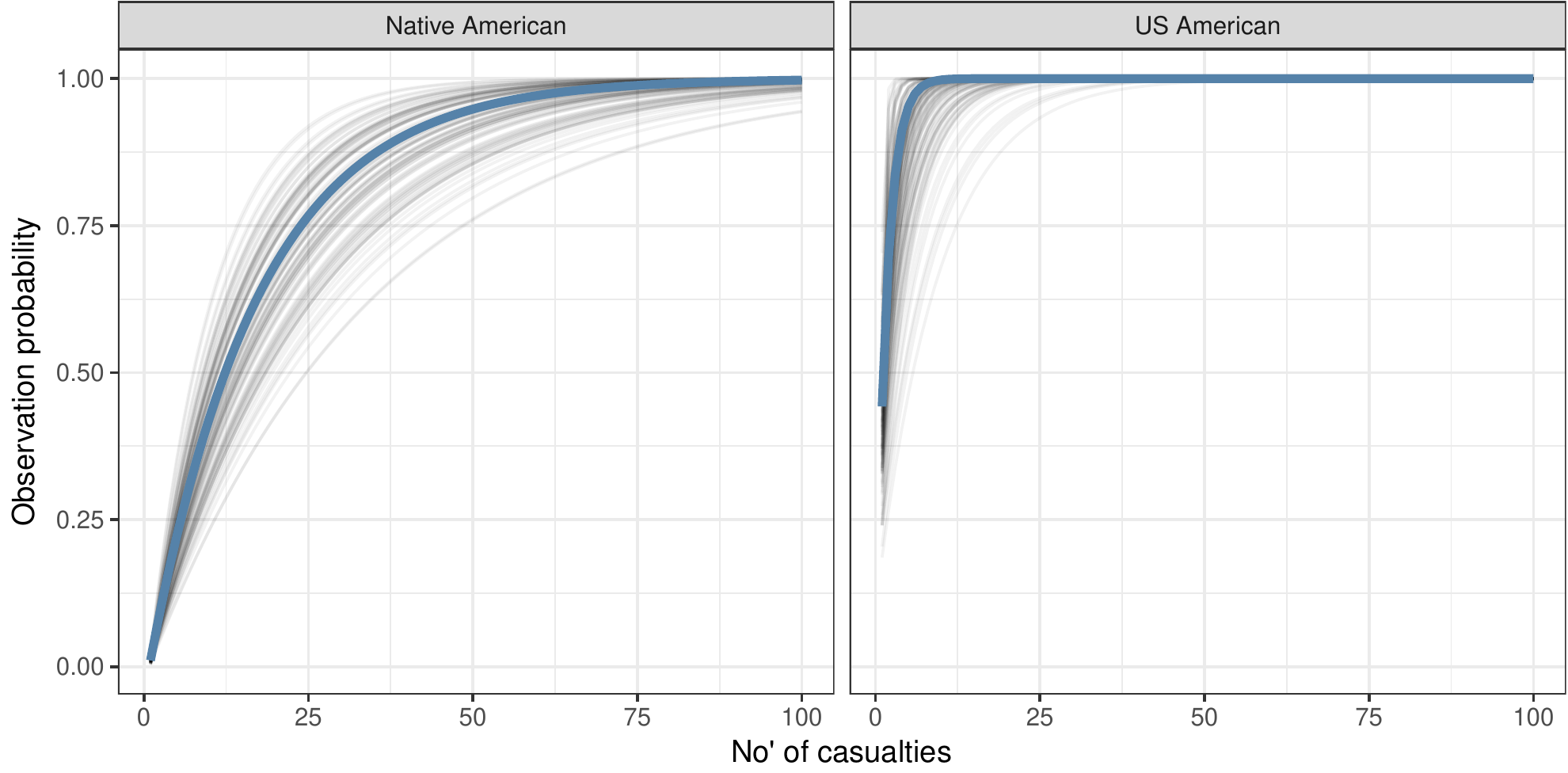}
  \caption{The probability of observing an event. The blue lines give the
    posterior mean estimate for the observation line. The light grey lines are
    $100$ samples from the posterior.}\label{F6}
\end{figure}

The parameters modeling the probability of observing an event considerably
differ between the two groups. Figure \ref{F6} plots the probability of
recording an event. One hundred samples from the posterior have been plotted, in
addition to the average probability. The posterior mean probability of observing
an event of size $1$ is $E(1-e^{-\lambda} \vert z) \simeq 0.010$, for the Native
Americans, but is significantly larger for the US Americans ($0.46$); where $z$
are the observed battles. The parameter governing the rate at which we perfectly
record events, $\mu$, is also larger for the US forces. Related, the posterior
mean estimate of $n_{true}$, which is $2,287$ and $20,551$ for the US and Native
American forces respectively. For comparison, $n_{obs}$ was $1,232$ and $1,297$.

We can obtain a quasi-estimate of $\xmin$, that is, the point where we are
unlikely to miss a battle, by calculating
\[
x_{0.95} = \arg \min_x (E(1 - \exp\{-\lambda - \mu (x-1)\}\vert z) > 0.95) \;,
\]
where the posterior expectation is calculated using samples from the posterior.
This gives point estimates of $x_{0.95} = 10$ and $x_{0.95} = 57$ for the US and
Native American forces, respectively.

As might be expected, rounding or heaping is more prevalent for the Native
American forces, with the probability of rounding almost twice as large as for
the US American forces (see figure \ref{F5} (g, h)).

\begin{figure}[t]
  \centering 
  \includegraphics[width=\textwidth]{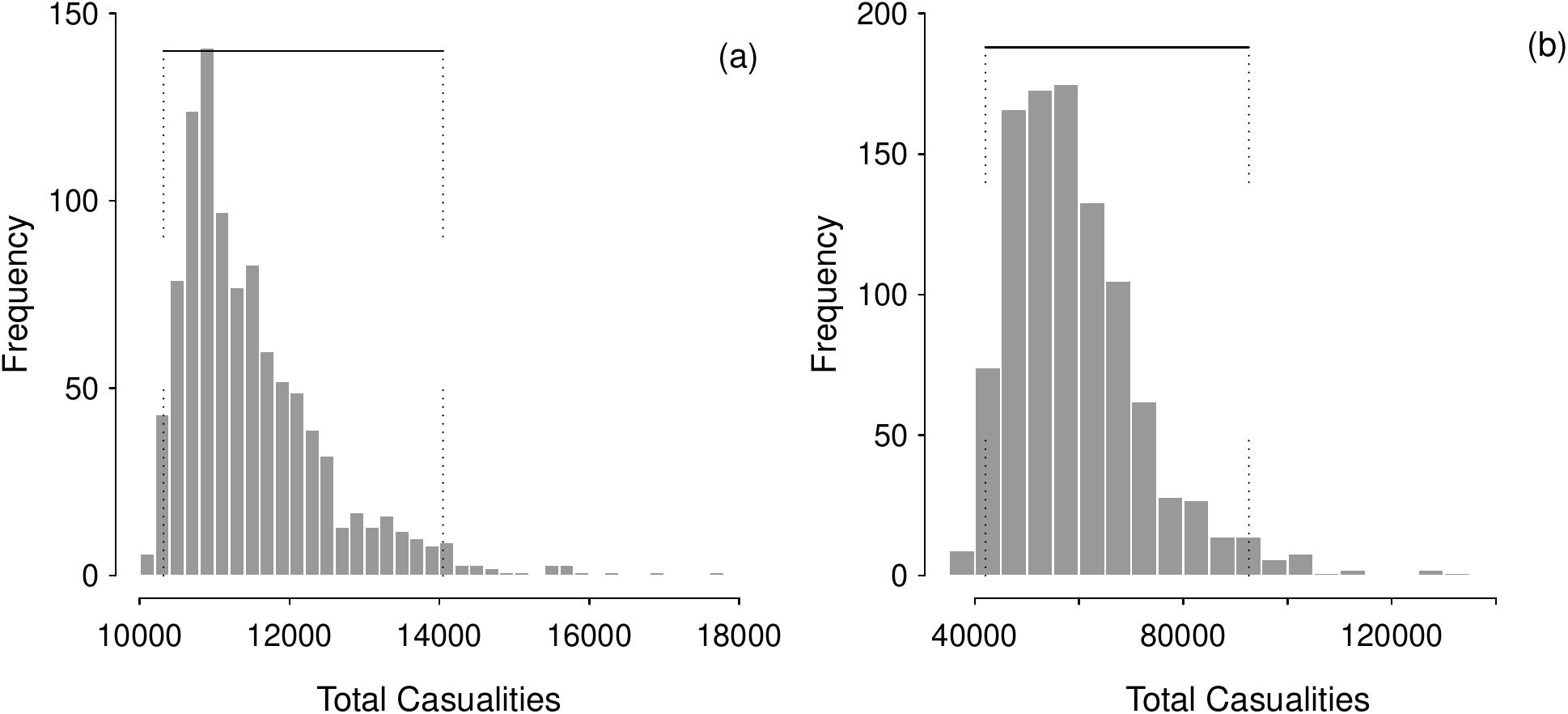}
  \caption{Predictive distributions for the observed and total number of
    casualties for the US and Native American forces, respectively. For each
    distribution, a 95\% prediction interval is shown.}\label{F7}
\end{figure}

We can use samples from the posterior distribution to infer the total number of
casualties. Figure \ref{F7} shows the predicted number of total casualties
sustained by each side. Clearly there is a more uncertainty in the Native
American casualties (as also demonstrated in figure \ref{F5}), with the mean
number of casualties being around $11,500$ and $60,000$. These numbers are
larger than the corresponding power law analysis of \cite{Friedman2013}, who
estimated $12,000$ and $53,000$ respectively. The reason for the difference is
two fold. First, Friedman ignored other sources of error; in particular rounding
which is significant in the Native American forces. Second, when estimating
uncertainty in the parameters, Friedman used a bootstrap procedure, whereas in
this analysis, we condition on what has already been observed. The difference in
analysis is noticeable when comparing figure 3 of \cite{Friedman2013} with
figure \ref{F7}. In the bootstrap version, there is considerable distributional
mass for casualty estimates less than $40,000$. Some of the bootstrapped samples
also suggested that the casualties have been over-estimated, which seems highly
unlikely.

\subsection{Model fit and sensitivity}\label{modelfit}

As with any Bayesian analysis, it is important to assess the sensitivity of the
posterior to the prior specification. Although the prior on the power law
coefficient was bounded (this study used a $U(1.5, 3)$ prior) the maximum
accepted value during the MCMC algorithm was less than $2.5$. Similarly, the
prior for the heaping coefficient $p$ was flat. The parameters governing the
missing observations did contain more information. However, switching to uniform
priors on $\mu$ and $\lambda$ did not substantially effect casualty inferences,
but did make tuning the MCMC algorithm more difficult.

A different functional form for the observation probability could also be used.
We investigated a quadratic form
\[
1 - \exp\{-\lambda - \mu(x-1) - \eta(x-1)^2\}
\] 
and a logistic function
\[
\frac{1}{1 + \exp(-\mu x)}\;.
\]
For each of these functions, the overall conclusions were similar with
inferences regarding $\alpha$ and $p$ being relatively unaffected.

\begin{figure}[!t]
  \centering 
  \includegraphics[width=0.95\textwidth]{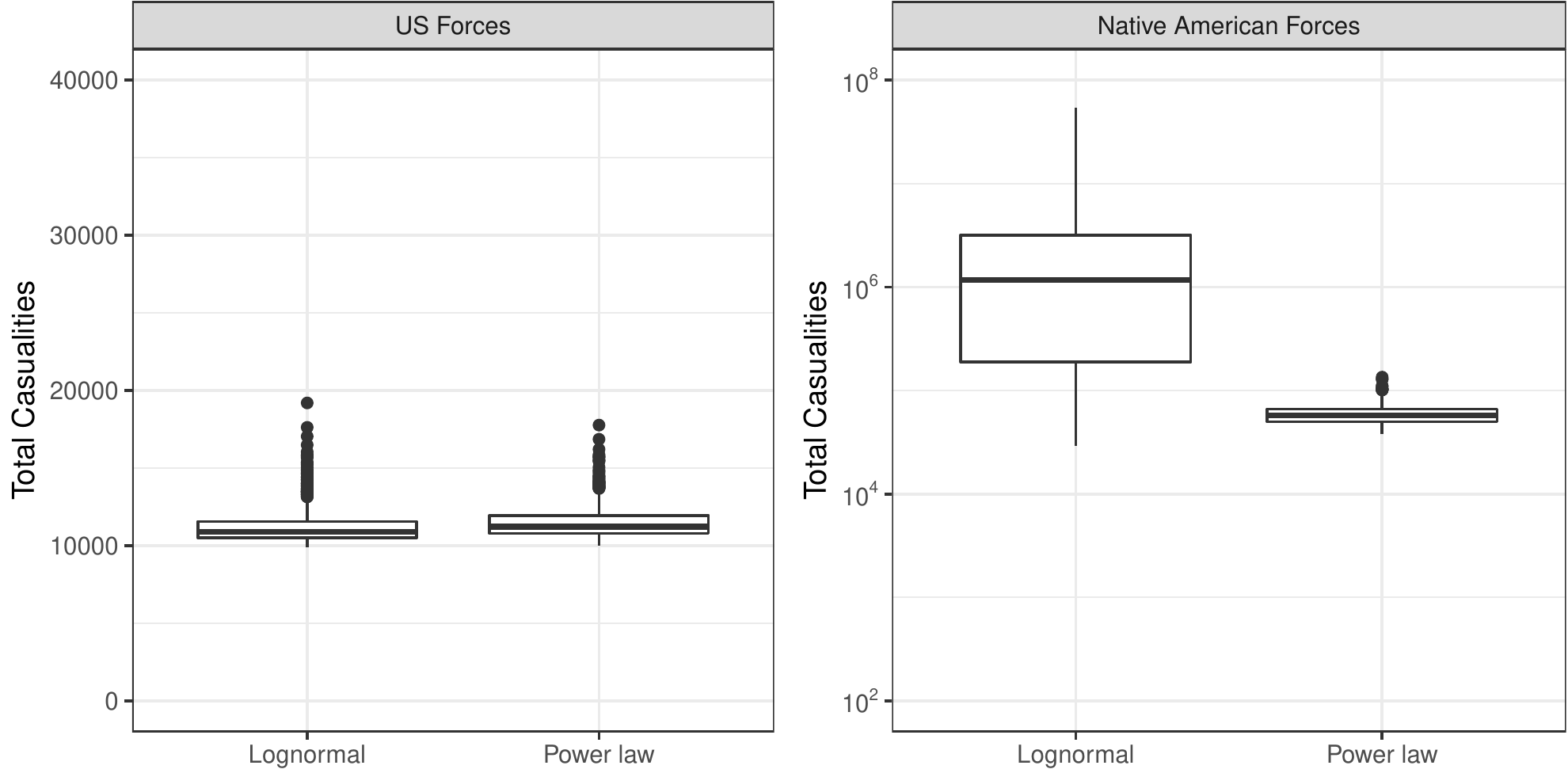}
  \caption{Predictive distributions for the number of casualties for the US and
    Native American forces, under the log normal and power law models. The power
    law data is the same as figure \ref{F6}.Note the change of scale in the
    Native American casualties.}
  \label{F8}
\end{figure}

To capture data rounding, the model only considers multiplies of five. However,
examining the Native American casualties, the two largest events were $1,000$
and $2,000$ individuals. It therefore seems likely that for larger events,
rounding was occurring to the nearest $100$ or $1,000$. However, there are few
large events occurring and so we decided against modelling this, and just note
that the overall estimates are unlikely to be affected by this omission.

Perhaps the strongest assumption that we made was assuming that the true
underlying distribution was a power law. This assumption was based on the
distribution of more recent conflicts. To assess the strength of this
assumption, we investigated how the prediction of the total number of casualties
altered if we assumed a (discrete) log normal distribution. Using uniform priors
for the log normal parameters and the same MCMC scheme as described previously,
we obtained the predictions given in figure \ref{F8}. The predicted number of
casualties for the US forces are broadly similar to the results in figure
\ref{F7}. The Native American forces has many more extreme points in the log
normal analysis, with the median number of predicted casualties increasing from
$66$ thousand to around a million casualties. However, the extreme results are
due to the perhaps unreasonable flat priors used for the log normal priors.
Interestingly, the estimate of the rounding parameters for each force are
relatively unaffected when switching to a log normal distribution.

\section{Discussion}

In many disparate research areas, underlying processes may generate events on
different orders of magnitude. In particular, since the system operates at
different levels, modelling the entire mechanism is difficult and so researchers
focus on the tail of the distribution. However, by purely focusing on the tail
region, it becomes more difficult to incorporate an error model.

This article builds on the work of \cite{Friedman2013} who used the power law
structure for prediction. However, by estimating directly $\xmin$, it made
extending the analysis more complicated when considering more realistic error
structures.

The American Indian war play a central role in the history of the United
states. However, due to missing data it has been difficult to quantify the
number lives lost during this time period. This article provides estimates for
the number casualties suffered by both sides. We estimate that the US forces
suffered around $12,000$ casualties in this conflict. As \cite{Friedman2013}
notes, this is approximately equal to the combined totals of the War of
Independence, the War of 1812, the Mexican-American War, and the Spanish-
American War. The Native Americans suffered far greater losses, around $60,000$.
Since the Native American population was around $400,000$ at the start of the
conflict, the number of casualties was catastrophic. To put this number into
context, as crude average, suppose the casualties are distributed equally
throughout the $115$ year conflict. This results in approximately $0.15\%$ of
total population dying in the conflict \textit{each year}. This is an order of
magnitude more than the United States lost in World War 2.

In this paper we modelled the available data and accounted for the different
sources of uncertainty. While this resulted in a more complex analysis, it also
yields more detailed insights, such as the amount of rounding in each data set.
Of course as \cite{Friedman2013} points out, we do not need power laws or
sophisticated statistics to establish that the American Indian war were
catastrophic for the Native Americans. Indeed, it has been suggested that up to
20 million Native Americans died as result of disease \citep{Pinker2011}. This
analysis attempts to better quantify the number of casualties associated with
armed conflicts.

The problem tackled in this paper does not provide a definite answer. Instead,
it relies heavily on expert opinion and insight. By building a more structured
model and using the Bayesian paradigm, we are able to channel prior beliefs
about the probability of observing events and the structure of the underlying
model into a predictive framework. The techniques described in this paper could
be applied to more recent conflicts, such as Iraq. Where it is difficult to
assess number of casualties sustained due to missing data.

The salient, but obvious, point raised in this paper, is that we rarely observe
data without error. The error structure could be as simple as Normal
perturbations to the true process, or something more complex as described in
this paper. Regardless, it is important to consider the impact on our inferences
if we ignore the underlying error structure. Indeed, many of the examples
considered in the original CSN paper have clearly been observed with error. By
switching to a Bayesian analysis, we have been able to properly account for the
different sources of error.

\section*{Computing details}

All simulations were performed on a machine with 4GB of RAM and with an Intel
quad-core CPU using R \citep{RCoreTeam2013}. The CSN power law fits were
obtained using the \texttt{poweRlaw} package \citep{Gillespie2015}. All code
associated with this paper can be obtained from
\begin{center}
https://github.com/csgillespie/plbayes
\end{center}

\section*{Acknowledgements}

We would like to thank Jeff Friedman and Richard Boys for their helpful comments
on this manuscript.

\bibliography{refs}

\newpage
\appendix

\clearpage
\end{document}